\documentclass[lettersize,journal]{IEEEtran}
\usepackage{amsmath,amsfonts}
\usepackage{algorithmic}
\usepackage{algorithm}
\usepackage{array}
\usepackage[caption=false,font=normalsize,labelfont=sf,textfont=sf]{subfig}
\usepackage{textcomp}
\usepackage{stfloats}
\usepackage{url}
\usepackage{verbatim}
\usepackage{graphicx}
\usepackage{cite}
\usepackage{times}
\usepackage{epsfig}
\usepackage{amsmath}
\usepackage{amssymb}
\usepackage{bm}
\usepackage{multirow}
\usepackage{balance}

\begin{document}

\title{RFGAN: RF-Based Human Synthesis}

\author{Cong~Yu,
	Zhi~Wu,
	Dongheng~Zhang,
	Zhi~Lu,
	Yang~Hu,\\
	and~Yan~Chen,~\IEEEmembership{Senior Member,~IEEE}
\thanks{C. Yu and Z. Lu are with University of Electronic Science and Technology of China. E-mail: congyu@std.uestc.edu.cn, zhilu@std.uestc.edu.cn}
\thanks{Z. Wu, D. Zhang, Y. Hu and Y. Chen are with University of Science and Technology of China, and Y. Chen is the corresponding author. E-mail: wzwyyx@mail.ustc.edu.cn, dongheng@ustc.edu.cn, eeyhu@ustc.edu.cn, eecyan@ustc.edu.cn}}
\maketitle

\begin{abstract}
This paper demonstrates human synthesis based on the Radio Frequency (RF) signals, which leverages the fact that RF signals can record human movements with the signal reflections off the human body. Different from existing RF sensing works that can only perceive humans roughly, this paper aims to generate fine-grained optical human images by introducing a novel cross-modal RFGAN model. 
Specifically, we first build a radio system equipped with horizontal and vertical antenna arrays to transceive RF signals.
Since the reflected  RF signals are processed as obscure signal projection heatmaps on the horizontal and vertical planes, we design a RF-Extractor with RNN in RFGAN for RF heatmap encoding and combining to obtain the human activity information. Then we inject the information extracted by the RF-Extractor and RNN as the condition into GAN using the proposed RF-based adaptive normalizations. Finally, we train the whole model in an end-to-end manner. To evaluate our proposed model, we create two cross-modal datasets (\textit{RF-Walk} \& \textit{RF-Activity}) that contain thousands of optical human activity frames and corresponding RF signals. Experimental results show that the RFGAN can generate target human activity frames using RF signals. To the best of our knowledge, this is the first work to generate optical images based on RF signals.
\end{abstract}


\section{Introduction}
\IEEEPARstart{V}{arious} recent works have built Radio Frequency (RF) sensing systems to perceive and understand the activities of humans. Compared with alternative sensing methods, RF sensing has improved usability due to the characteristics of RF signals, for example, the RF signals can work in all-day and all-weather scenarios, the sensing is contactless etc.. Existing RF-based human sensing works mainly include human position tracking~\cite{adib2013see, kotaru2015spotfi, li2017indotrack, ghazalian2020energy, qian2017widar, chen2019residual, zhang2018multitarget, zhang2020mtrack, liu2021multi}, human speed estimation~\cite{chen2020speednet, qian2018enabling, wu2015non}, human keypoint prediction~\cite{zhao2018through, zhao2018rf, wang2019person, jiang2020towards}, and gesture recognition~\cite{niu2021understanding}. However, these works can only perceive humans roughly and the sensing results usually lack fine details and are not as intuitive as optical sensing results.

\begin{figure}
	\begin{center}
		\includegraphics[width=\columnwidth]{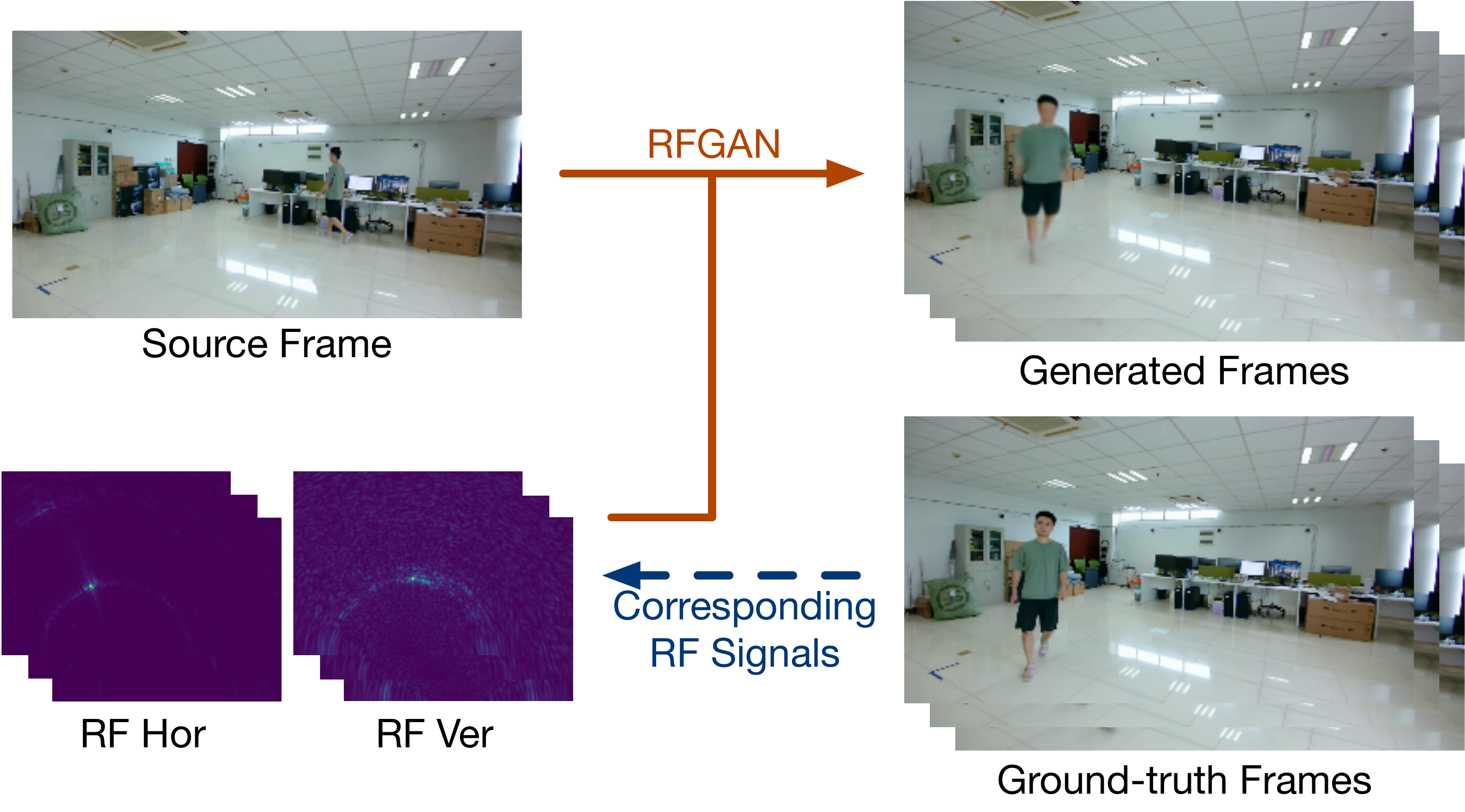}
	\end{center}
	\caption{With a source frame as reference, the RFGAN model can synthesize the target human activities based on the RF signals.}
	\label{fig:intror}
\end{figure}

In recent years, Generative Adversarial Networks (GAN)~\cite{goodfellow2014generative} have achieved promising results in modeling complex multimodal data and synthesizing realistic images. 
Furthermore, to generate meaningful images that meet actual requirements, many conditional GAN models have been proposed to control the generated results. Researchers have explored various kinds of conditions, e.g., category labels~\cite{ConditionalarXiv14}, text descriptions~\cite{GenerativeICML16, StackGANICCV17, li2020manigan}, and images~\cite{Image-to-ImageCVPR17, CartoonGANCVPR18, tang2019cycle, park2019semantic, PersonalizedFashionICCV19, chen2019quality, zhang2020supervised, emami2020spa, richardson2021encoding}. From technology perspective, most existing GAN models require the conditions to be able to guide the GAN model explicitly~\cite{ConditionalarXiv14, Image-to-ImageCVPR17, tang2019cycle, park2019semantic, richardson2021encoding} or can be transformed to conditional variables for GAN using an existing pre-trained model~\cite{GenerativeICML16, StackGANICCV17, li2020manigan}.

In this paper, we propose a solution to overcome the limitation of existing RF-based human activity sensing by making the results more visually intuitive, which is valuable in practice. We leverage the power of GAN models to generate photo-realistic sensing results from RF signals. Specifically, a photograph of the people in the scene is provided so that the GAN model has sufficient information about the visual appearance of the people and the environment of the scene. 
We use millimeter wave (mmWave) radars to build our radio system, which is equipped with two antenna arrays, horizontal and vertical ones, to obtain the RF signals that reflect off the human body. We process the horizontal and vertical RF signal reflections to horizontal and vertical RF heatmaps, which record the activity information of the human. 
The RF signal is a new kind of conditional data for GAN models. Due to the characteristics of RF signals, the resolution of the horizontal and the vertical RF heatmaps are relatively low. Besides, their spatial structures are quite different from optical images. Therefore, to utilize the RF signals as the conditional data to guide the GAN model, some challenges need to be addressed: firstly, we need to train the RF conditioning encoding network without supervision labels to obtain the desirable human activity information; secondly, the information from the horizontal and vertical RF heatmaps need to be fused to characterize the overall human activity; thirdly, the fused information needs to be injected into the GAN model properly.

To tackle the above challenges, we design dual RF-Extractors and RNNs in the RFGAN model, one in the generative part and the other in the discriminative part, and we train them by adversarial learning. 
Two CNN encoders in the RF-Extractor are used to extract features from the horizontal and vertical RF heatmaps, respectively. Then a novel fusion operation is designed to fuse the information by building relationships between the extracted features. 
To inject the fused information into the GAN model, inspired by~\cite{huang2017arbitrary, Karras_2019_CVPR, park2019semantic}, we propose to modify the distributions of the latent features in GAN by using a RF-based adaptive normalization. 
Furthermore, we create two cross-modal datasets (\textit{RF-Walk} \& \textit{RF-Activity}) that consist of optical human activity frames and corresponding RF signals to train and test our proposed RFGAN model. The experimental results show that the RFGAN can generate better human results than alternative methods. 

Since the RF signals do not rely on visible lights and can traverse occlusions, our proposed RFGAN model can also work when lights dim or the human is occluded by barriers. For example, when the environment is favorable, we capture a human frame as the source. Our radio system can record the RF signal reflections when the illumination becomes bad or the human is in occlusions. The proposed RFGAN model can synthesize human activities based on these collected multimodal data.

Therefore, the main contributions of this paper can be summarized as follows:
\begin{itemize}
	\item [1.] We propose a novel RFGAN model to enable RF-based human synthesis. To the best of our knowledge, this is the first work to generate human images from the mmWave radar signals. There are many potential applications that can be derived from this task, e.g., fine-grained human perception and all-day monitoring systems in the smart home.
	\item [2.] Technically, for the new kind of conditional data, i.e., the RF signals, we propose to train the RF conditioning encoding network, i.e., the RF-Extractor and RNN, by adversarial learning. Then we design a novel fusion operation to fuse the horizontal and vertical RF information, which is an effective approach for overall human activity sensing from the two-dimensional RF heatmaps. Due to the spatial structure difference, we propose to use the RF-based adaptive normalizations to inject the fused information into the GAN model.
	\item [3.] We create two cross-modal datasets, i.e., \textit{RF-Walk} and \textit{RF-Activity}, which contain thousands of optical human activity frames and corresponding RF signals. The datasets will be released to public.
\end{itemize}

\section{Related Work}
\noindent\textbf{Conditional GAN}
Many research works have shown that GAN~\cite{goodfellow2014generative} has the capability of generating realistic images based on the given conditional data. For example, \cite{ConditionalarXiv14} utilize category labels to generate target digit images. Some works~\cite{Image-to-ImageCVPR17, CartoonGANCVPR18, tang2019cycle, park2019semantic, PersonalizedFashionICCV19, chen2019quality, zhang2020supervised, emami2020spa, richardson2021encoding} introduce the GAN-based image-to-image translation frameworks. For some more complex conditional data, e.g., text data, \cite{GenerativeICML16, StackGANICCV17, li2020manigan} use existing pre-trained models to transform the text into conditioning variables for GAN.
To employ these conditions in the networks, some works~\cite{huang2017arbitrary, Karras_2019_CVPR, park2019semantic} find that utilizing the conditional normalization in the hidden layers can contribute to generating target images. 
In our case, we take RF signals as the condition to guide the image synthesis, which is a new cross-modal conditional data that has obscure guidance for GAN and has no existing pre-trained model for conditioning encoding.

\noindent\textbf{RF-Based Human Perception}
Recent years have witnessed much interest in using RF signals to enable various human perception tasks~\cite{he2020wifi}, including indoor localization and tracking~\cite{adib2013see, kotaru2015spotfi, ghazalian2020energy, li2017indotrack, qian2017widar, zhang2018multitarget, zhang2020mtrack, liu2021multi}, human speed estimation or human movement detection~\cite{chen2020speednet, qian2018enabling, wu2015non, niu2021understanding}, human identification~\cite{zeng2016wiwho, fan2020learning, hsu2019enabling}, and human vital signs inference~\cite{zhang2019breathtrack, yue2018extracting, rahman2015dopplesleep, zhang2019sj, hsu2017zero}.
Besides the above signal-processing-based methods, approaches based on deep learning are also utilized to handle radio human perception. For example, \cite{zhao2017learning} combines convolutional and recurrent neural networks to learn sleep stages from radio signals. \cite{zhao2018through, zhao2018rf} propose to predict the 2D/3D human keypoints based on RF signals by building a teacher-student network model. 
In this paper, we propose to use RF signals for human image synthesis by combining conditional GAN models. 

\begin{figure}
	\begin{center}
		\includegraphics[width=0.96\columnwidth]{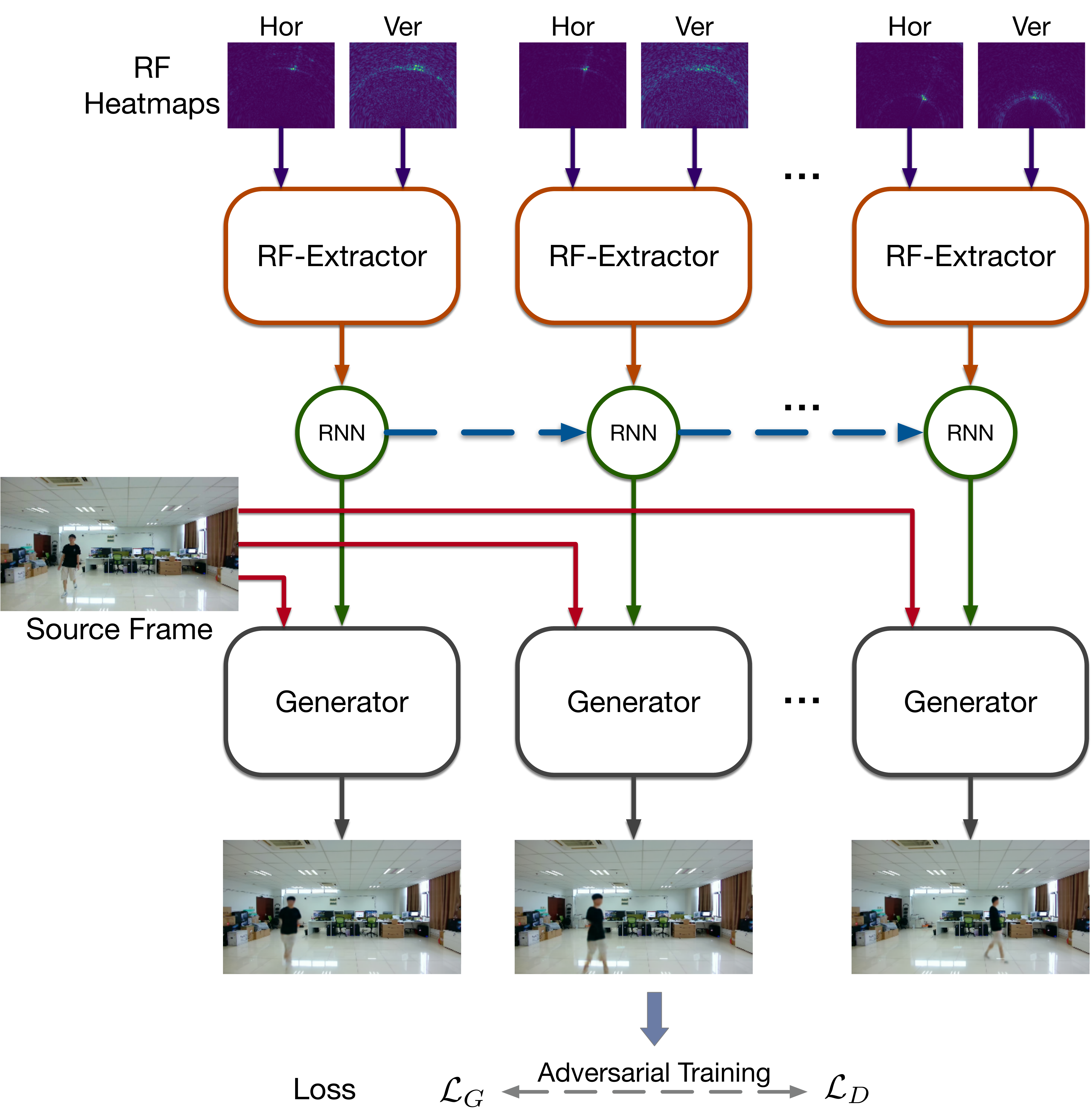}
	\end{center}
	\caption{The architecture of the RFGAN model for generating sequential human activity frames.}
	\label{fig:model}
\end{figure}

\begin{figure*}
	\begin{center}
		\includegraphics[width=0.95\textwidth]{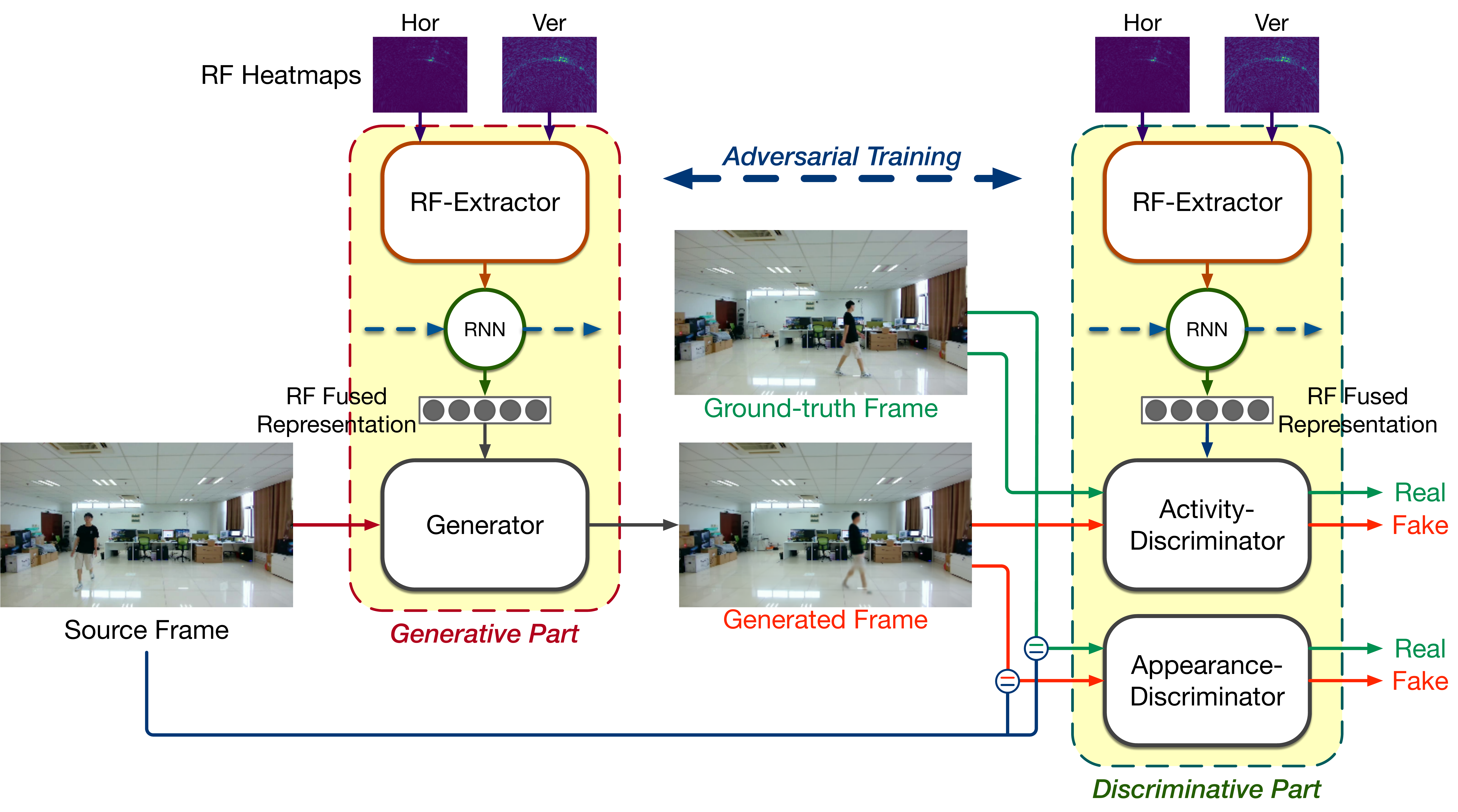}
	\end{center}
	\caption{The training framework of RFGAN at one moment. It consists of a generative part and a discriminative part. The whole model is trained by adversarial learning in an end-to-end manner.}
	\label{fig:rfgan}
\end{figure*}

\noindent\textbf{Sequence Modeling} 
Recurrent neural networks, such as vanilla RNN, GRU and LSTM, have been widely used for processing sequential data, such as text and speech. They have also been successfully applied to model the temporal dependencies in videos for various vision problems, such as video classification~\cite{yue2015beyond}, action recognition~\cite{song2017end, zhang2018fusing, si2018skeleton, agethen2019deep, si2019attention},  object segmentation~\cite{ventura2019rvos}, video prediction~\cite{lu2017flexible}, etc.  In this work, considering that the RF signals are sequential data and the RF heatmaps are the samples at different moments, we utilize recurrent neural networks as the backbone of our model to perceive human activities from RF signals and synthesize corresponding optical images.


\section{Preliminary}
Our method relies on transmitting RF signals and receiving the reflections. We adopt Frequency Modulated Continuous Wave (FMCW) and linear antenna arrays for signal transceiving. Inspired by ~\cite{zhao2018through}, our radio system is equipped with two antenna arrays: horizontal and vertical ones, which are utilized to acquire the signal projections on the plane parallel to the ground and the plane perpendicular to the ground, respectively. 
Hence, the RF data is composed of both horizontal and vertical heatmaps.

Compared with camera-based visual data, RF signals have some different characteristics. Firstly, RF signals have much lower resolution. The resolution is determined by the bandwidth of the signal and the aperture of the antenna array~\cite{richards2014fundamentals}. In our system, the depth resolution is about 7.5cm, and the angular resolution is about 1.3 degrees. Secondly, the RF signals suffer from severe multi-path propagation in an indoor environment~\cite{zhang2020mtrack}, which introduces severe interference in the received signals. Thirdly, the RF signals have different representations of the scene compared with the camera, i.e., horizontal and vertical projections.

\section{RFGAN}

The RFGAN model aims to generate sequential human activity frames using a sequence of RF heatmaps (horizontal \& vertical) and a source frame. 
To extract and combine the human activity information from the horizontal and vertical RF heatmaps, we design a RF-Extractor, which is built with a sequence model, i.e., the Recurrent Neural Network (RNN), to process the RF sequence.
To generate optical human activity frames, we utilize the Generative Adversarial Network (GAN) as the main technological approach in our model, where the source frame is fed as the input layer and the information extracted from RF heatmaps is the condition of GAN. 

The architecture of RFGAN model is shown in Figure~\ref{fig:model}. The RNN is the backbone of the model, which is designed for sequence data processing and generation. The RF-Extractor and the Generator are plugged into both sides of the RNN to process RF heatmaps and generate human frames. 
In the following subsections,  we first introduce the training framework of the model and then discuss the network structures of the RF-Extractor, the RNN, and the RF-based Generator and Discriminators in detail. Finally, we describe the loss functions used to train the whole model. 

\subsection{Training Framework}
The proposed human synthesis model aims to generate sequential human frames from a source frame and the corresponding sequential RF heatmaps. Figure~\ref{fig:rfgan} shows the adversarial training framework of the human synthesis model at one moment, which consists of a generative part and a discriminative part. The generative part contains a RF-Extractor, a RNN, and a Generator. The RF-Extractor and RNN extract the human position and posture information from the corresponding RF heatmaps and represent it as a RF fused representation. For the Generator, the source frame is fed as the input layer, and the extracted RF fused representation controls the network through normalization at the convolution layers. The output is the generated human frame. There are two Discriminators in the discriminative part. The Activity-Discriminator is designed to ensure that the human position and posture in the generated frame are consistent with the RF signal. It takes the generated frame as the input layer. The RF fused representation extracted by the RF-Extractor and RNN in the discriminative part is used as condition of this Discriminator. The Appearance-Discriminator ensures that the generated frame maintains the same visual information, such as human appearance, with the source frame, thus the generated frame is concatenated with the source frame at the input layer.

Note that the RF-Extractors and RNNs in the generative part and the discriminative part have the same network structures and RF inputs, but they do not share the network parameters. 
In the previous GAN literature, the conditional variables that fed into the Generator and the Discriminator are obtained by the same network model, mainly due to the existence of the pre-trained model to enable the desirable conditioning encoding. However, in our task, there is no existing pre-trained model for RF encoding. Thus, we propose dual RF-Extractors and RNNs under an adversarial training framework to learn to transform the RF heatmaps, one for the generation task and the other for the discrimination task. 
Specifically, the RF-Extractor and RNN in the generative part update with the Generator, whereas the RF-Extractor and RNN in the discriminative part update with the Activity-Discriminator. The update process is adversarial training and the whole model is trained in an end-to-end manner.

\begin{figure}
	\begin{center}
		\includegraphics[width=0.93\columnwidth]{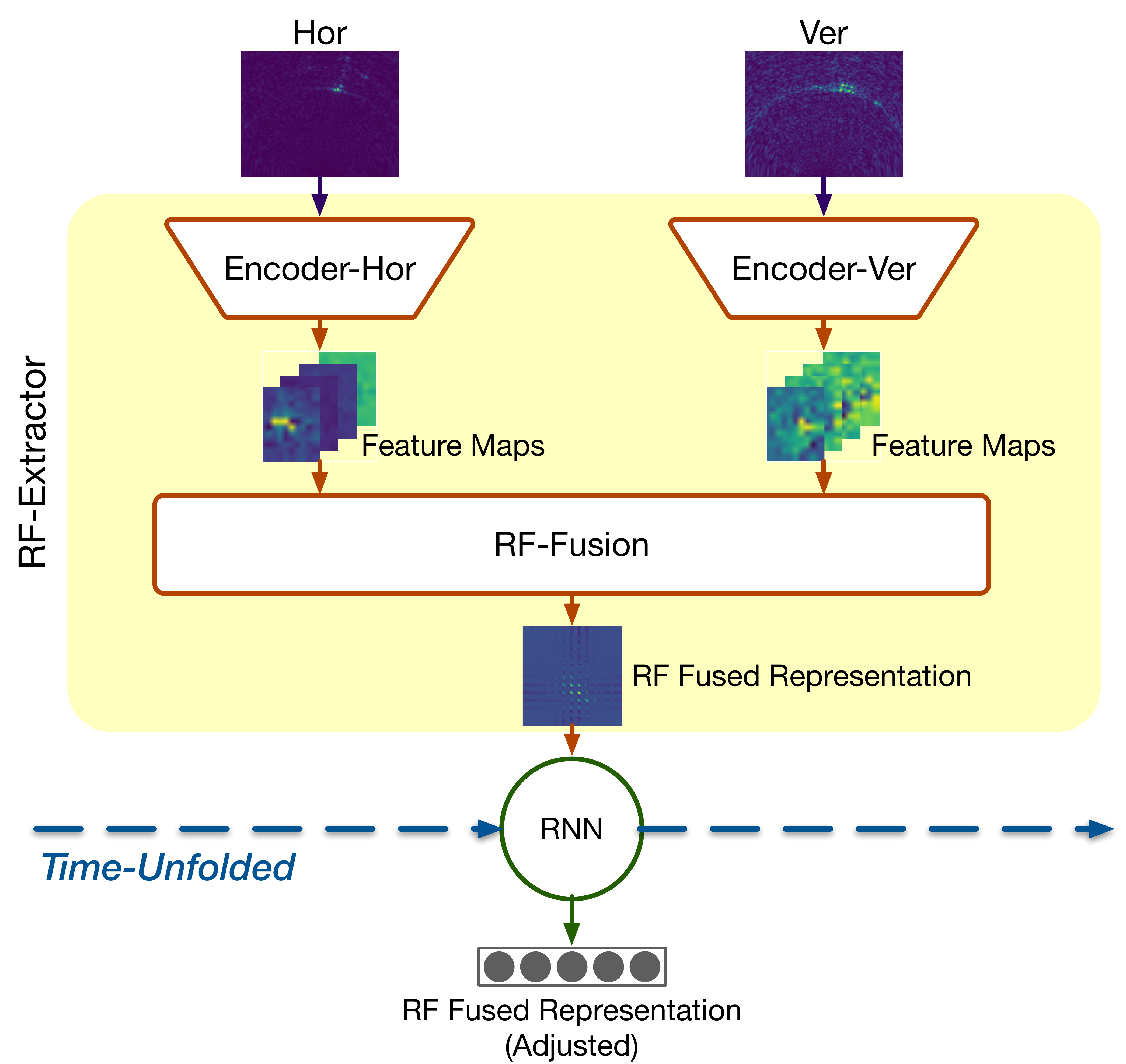}
	\end{center}
	\caption{The structure of RF-Extractor. It consists of two CNN encoders, a fusion operation, and a RNN.}
	\label{fig:rfextractor}
\end{figure}

\subsection{RF-Extractor \& RNN}
The horizontal RF heatmaps and the vertical RF heatmaps record human activities from different viewpoints and each of them only contains partial human activity information, i.e., the horizontal RF heatmap is a projection of the signal reflections on a plane parallel to the ground, which leads to the loss of the human height information, whereas the vertical heatmap is a projection of the reflected signals on a plane perpendicular to the ground and the human width information is missed. Thus, it is a challenge that how to extract and combine the horizontal and vertical RF information to get the whole human activity information.

\begin{figure}
	\begin{center}
		\includegraphics[width=0.93\columnwidth]{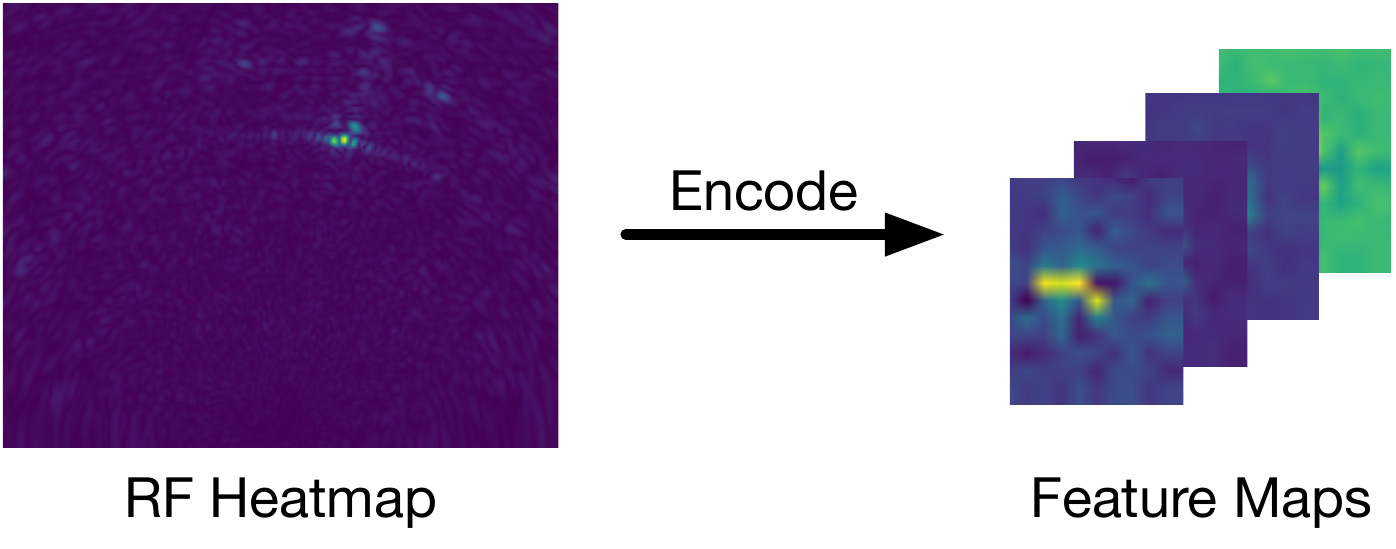}
	\end{center}
	\caption{The RF heatmap and the corresponding feature maps.}
	\label{fig:encode}
\end{figure}

In our proposed network structure, as shown in Figure~\ref{fig:rfextractor}, we first use two standard CNN encoders to transform the horizontal and vertical RF heatmaps into feature maps, respectively. 
The original RF heatmaps record the reflected signals throughout the whole room. After the differential operation along the time, only the signals introduced by the moving human are retained. As shown in left part of Figure~\ref{fig:encode}, we can find the signal reflections from the moving human (bright area) only occupy a very small area of the RF heatmap. Therefore, we use an encoder that consists of several convolution layers to reduce the RF heatmap size and focus on the bright area. Since the values of signal reflections from no human areas (dark areas) are very small and close to $0$, the convolution results, which are denoted as feature maps (shown in right part of Figure~\ref{fig:encode}), can capture the human posture information from the bright area and the human position information from the location of the bright area.

\begin{figure}[H]
	\begin{center}
		\includegraphics[width=0.96\columnwidth]{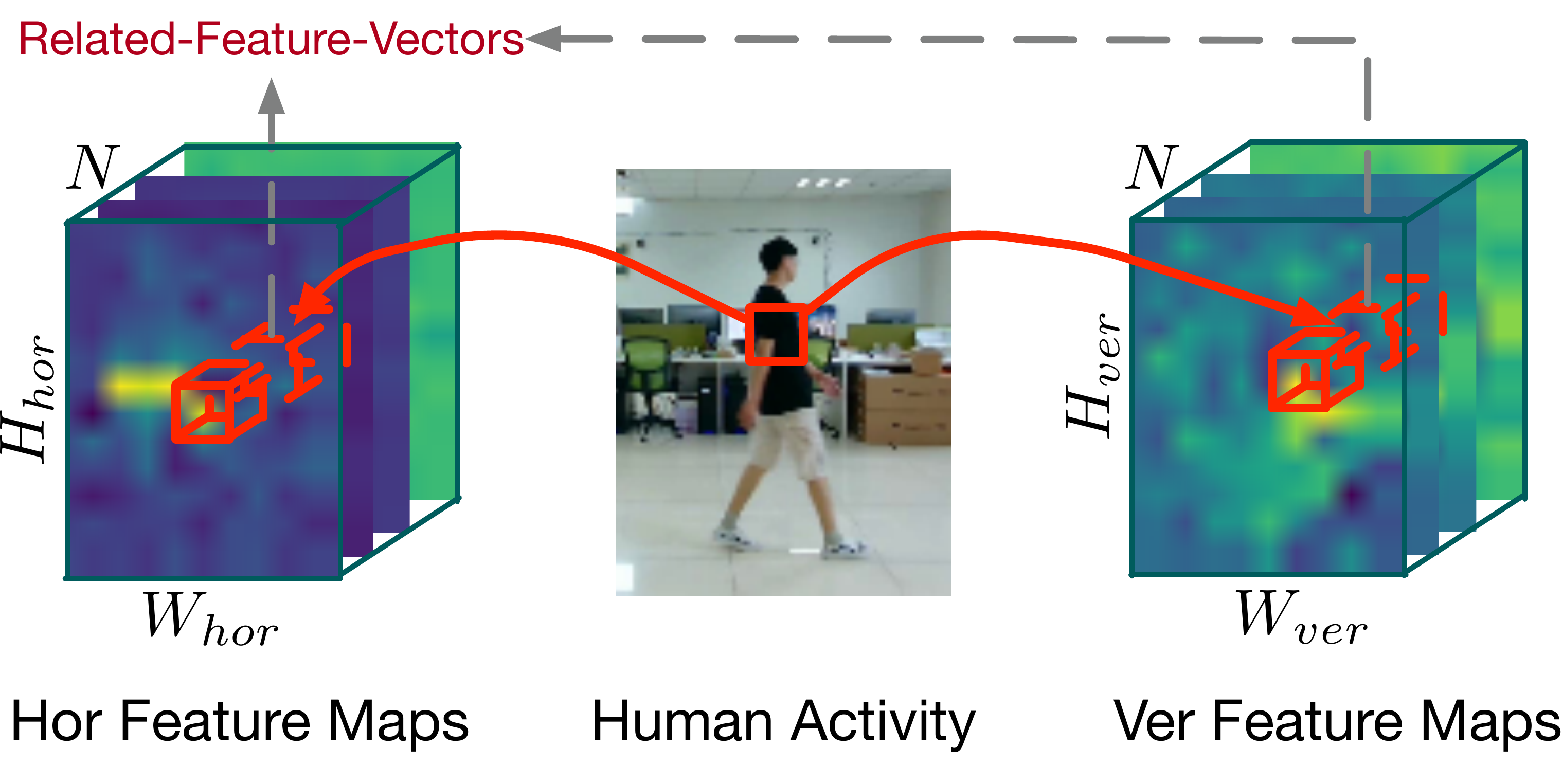}
	\end{center}
	\caption{The horizontal and vertical feature maps contain the human activity information on the horizontal and vertical plane, respectively. The red cuboids are the related-feature-vectors, which contain the activity information introduced by the same human body part.}
	\label{fig:relatedpixel}
\end{figure}

After encoding RF heatmaps, a fusion operation (RF-Fusion) is proposed to combine the horizontal and vertical feature maps into a fused representation.
As shown in Figure~\ref{fig:relatedpixel}, the horizontal feature maps can be represented as a $H_{hor} \times W_{hor} \times N$ tensor, which uses $H_{hor} \times W_{hor}$ feature vectors on a horizontal plane to record the human activity information, and each feature vector is $N$ dimensional. For the vertical feature maps, $H_{ver} \times W_{ver}$ feature vectors are used on a vertical plane to record the human activity information, and each feature vector is also $N$ dimensional. We refer to the feature vectors in the horizontal and vertical feature maps as related-feature-vectors if they record the activity information introduced by the same human body part (see Figure~\ref{fig:relatedpixel}). Combining these related-feature-vectors can help characterize the overall human activity. However, it is difficult to find the one-to-one correspondence between them directly.

\begin{figure}
	\begin{center}
		\includegraphics[width=\columnwidth]{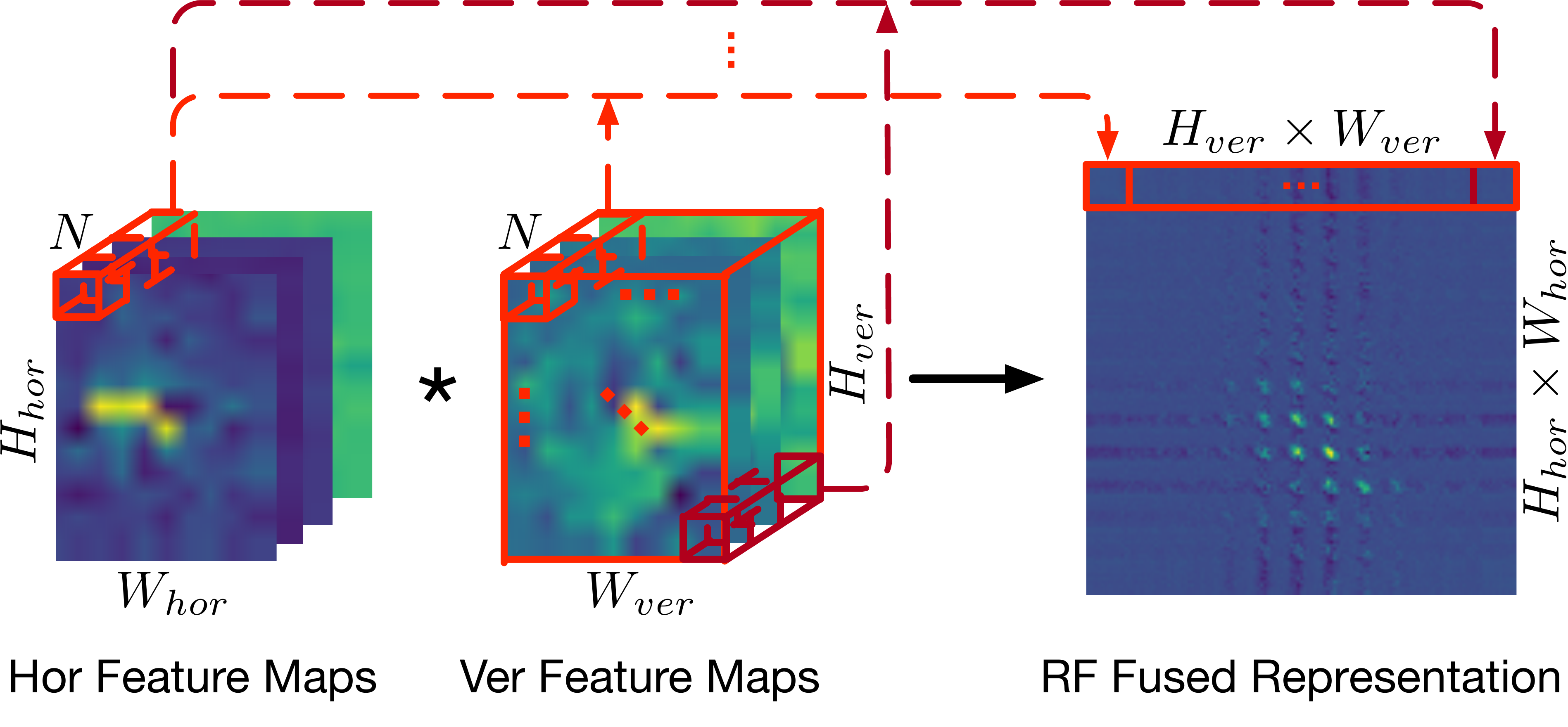}
	\end{center}
	\caption{The RF-Fusion operation.}
	\label{fig:rffusion}
\end{figure}

To address this problem and bridge these related-feature-vectors, we define RF-Fusion as follows: 
for each feature vector in the horizontal feature maps, the dot product is applied with every feature vector in the vertical feature maps, and the results are denoted as a RF fused representation. For example, as shown in Figure~\ref{fig:rffusion}, the dot products between the first horizontal feature vector and every vertical feature vector generate $H_{ver} \times W_{ver}$ values, which are the first row of the RF fused representation. In such a way, we can obtain the RF fused representation as follows:
\begin{equation}
	\begin{split}
		\label{eqn:fusion}
		R&(i, j)=\frac{H(i)V(j)^T}{\sqrt{N}}, \\ i \in [0, H_{hor}& \times W_{hor}), j \in [0, H_{ver} \times W_{ver}),
	\end{split}
\end{equation}
where $R(i, j)$ is the value at the point $(i, j)$ in the RF fused representation, $H(i)$ and $V(j)$ refer to the feature vector in the horizontal feature maps and the feature vector in the vertical feature maps. The denominator $\sqrt{N}$ is to scale the values.

\noindent\textbf{Why does RF-Fusion work?} The traditional feature map fusion approach is to concatenate feature maps along the channel directly, which is effective when feature maps have the same spatial structure, i.e., the feature vectors in the different feature maps are aligned and can be combined by concatenating. However, in the RF fusion step, for a given feature vector in the horizontal feature maps, we do not know which feature vector in the vertical feature maps is related to it. Thus our proposed RF-Fusion builds relationships between every feature vector in the horizontal feature maps and every feature vector in the vertical feature maps. 
For the learned RF-Extractor, the related-feature-vectors in the horizontal and vertical feature maps are supposed to be highly correlated and lead to generating large values in the RF fused representation, which are shown as bright points. Therefore, the distribution and values of these bright points can characterize the overall human activity.

Finally, the RF fused representations are fed into the RNN to get adjustments. We propose this procedure based on the following fact: human activities, such as arm swing, leg raising, etc., are generally continuous, thus the RF fused representation that contains the human position and posture information at a certain moment is interrelated with several preceding and subsequent RF fused representations. The RNN model can adjust the current representation by considering its neighbors, and the adjusted results would contain more smooth and more accurate human activity information. In our model, we use a three-layer BiLSTM as the proposed RNN, and the hidden states in the last layer are adopted as the adjusted RF fused representations.

\begin{figure}
	\begin{center}
		\includegraphics[width=\columnwidth]{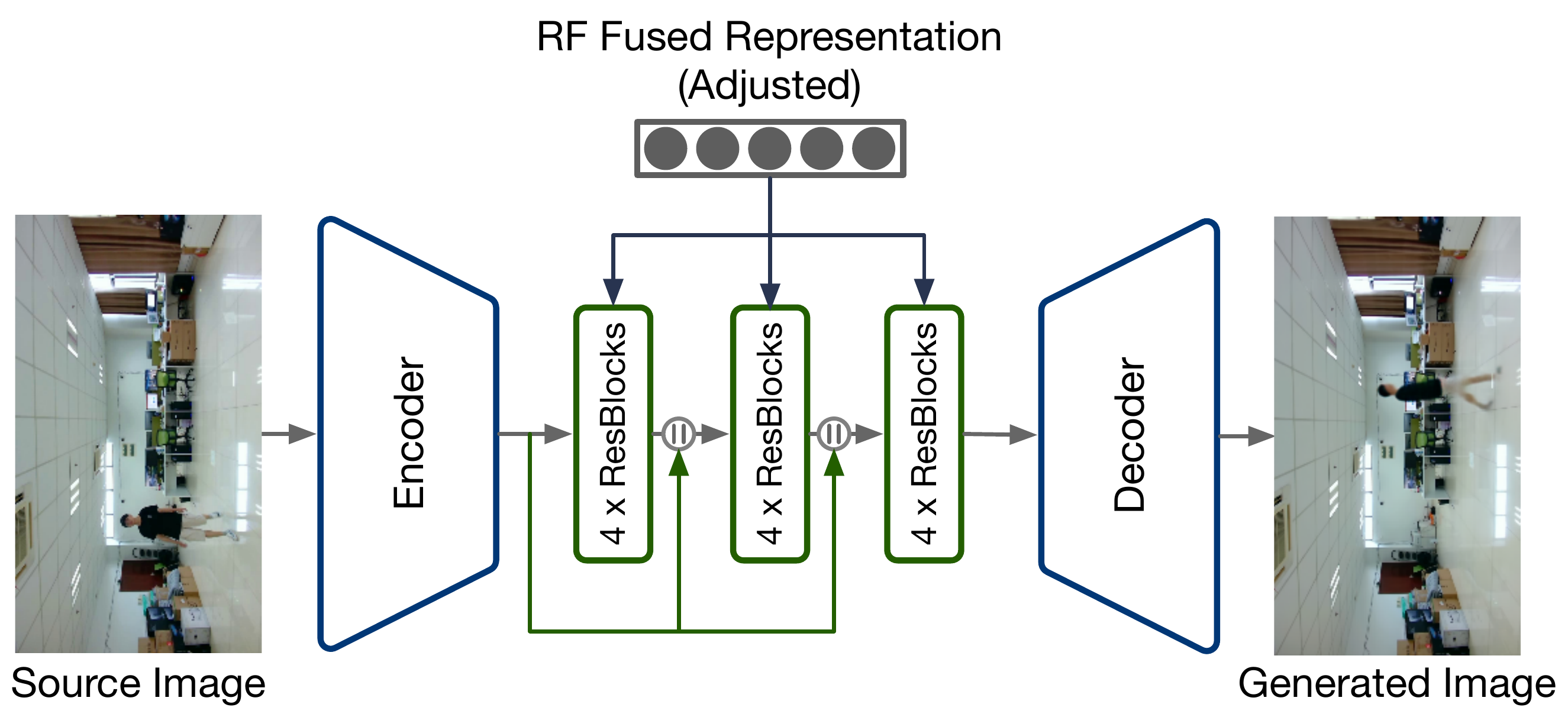}
	\end{center}
	\caption{The structure of Generator.}
	\label{fig:generator}
\end{figure}

\subsection{RF-Based Generator \& Discriminators}
The Generator in our model consists of an encoder, several residual blocks, and a decoder (see Figure~\ref{fig:generator}). The encoder and the decoder contain the same numbers of convolutional and deconvolutional layers. The residual blocks are divided into several groups and each group has the same numbers of blocks. The feature extracted from the source frame is concatenated with several feature maps in the residual blocks to maintain the appearance information. For the Activity-Discriminator and the Appearance-Discriminator, we use the network structures inspired by PatchGAN~\cite{Image-to-ImageCVPR17}. They both consist of convolutional layers, where the first layer does not use the normalization and the last layer is only a convolution to produce a 1-dimensional output.

Specifically, to enable the RF-based condition setting in RFGAN, we propose a RF-based adaptive instance normalization (RF-InNorm) in the hidden layers of Generator and Activity-Discriminator, which injects the RF fused representation by modifying the feature distribution. The RF-InNorm is defined as 
\begin{equation}
	\label{eqn:RFInNorm}
	\text{RF-InNorm}(\bm{f}^{n})=F_{\gamma}^n(\bm{h})\cdot\frac{\bm{f}^{n}-\bm{\mu}^n}{\bm{\sigma}^n}+F_{\beta}^n(\bm{h}),
\end{equation}
where $\bm{f}^n$ is the feature map of the $n$-th layer in the Generator or Activity-Discriminator, $\bm{\mu}^n$ and $\bm{\sigma}^n$ are the mean and standard deviation of the feature map.
$\bm{h}$ refers to the RF fused representation, $F_{\gamma}^n(\cdot)$ and $F_{\beta}^n(\cdot)$ are the learned nonlinear functions, which specialize $\bm{h}$ to RF-based modulation parameters.
Therefore, the feature map $\bm{f}^n$ is first normalized and then scaled and biased by $F_{\gamma}^n(\bm{h})$ and $F_{\beta}^n(\bm{h})$ to incorporate the RF fused representation condition. 

For the Appearance-Discriminator, the source frame condition is concatenated with the input and fed into the network. 

\subsection{Loss Functions}
The training process of the RFGAN is a two-player minimax game between the generative part and the discriminative part. For the discriminative part, we set the loss for the Activity-Discriminator and the RF-Extractor and RNN as:
\begin{equation}
	\begin{split}
		\label{eqn:d_act_loss}
		\mathcal{L}^{act} = \mathcal{L}^{act}_{LSD}+\lambda\mathcal{L}^{act}_{GP},
	\end{split}
\end{equation}
where $\mathcal{L}^{act}_{LSD}$ is the adversarial loss inspired by LSGAN~\cite{mao2017least}, $\mathcal{L}^{act}_{GP}$ is the gradient regularization term that penalizes the discriminator gradients only on the true data to stabilize the training process~\cite{mescheder2018training}, which can be calculated by
\begin{equation}
	\begin{split}
		\label{eqn:d_act_lsd}
		\mathcal{L}^{act}_{LSD}=&\mathbb{E}_{\bm{x}_r \sim \mathbb{P}}\left [(D_{act}(\bm{x}_r|E_{dis}(\bm{r}_h,\bm{r}_v))-1)^2\right ] + \\ &\mathbb{E}_{\bm{x}_f \sim \mathbb{Q}}\left [(D_{act}(\bm{x}_f|E_{dis}(\bm{r}_h,\bm{r}_v))-0)^2\right ],
	\end{split}
\end{equation}
and
\begin{equation}
	\begin{split}
		\label{eqn:d_act_gp}
		\mathcal{L}^{act}_{GP}= \mathbb{E}_{\bm{x}_r \sim \mathbb{P}} [\left\| \nabla D_{act}(\bm{x}_r|E_{dis}(\bm{r}_h,\bm{r}_v))\right\|_2^2],
	\end{split}
\end{equation}
where $D_{act}$ is the Activity-Discriminator, $E_{dis}$ is the RF-Extractor and RNN in the discriminative part, $\bm{x}_r$ and $\bm{x}_f$ are the ground-truth and the generated human frame, respectively, $\bm{r}_h$ and $\bm{r}_v$ refer to the horizontal RF heatmap and the vertical RF heatmap, respectively.

For the Appearance-Discriminator, the loss function is similar to the loss for the Activity-Discriminator and the RF-Extractor and RNN:
\begin{equation}
	\begin{split}
		\label{eqn:d_app_loss}
		\mathcal{L}^{app} = \mathcal{L}^{app}_{LSD}+\lambda\mathcal{L}^{app}_{GP},
	\end{split}
\end{equation}
the $\mathcal{L}^{app}_{LSD}$ and $\mathcal{L}^{app}_{GP}$ are calculated by
\begin{equation}
	\begin{split}
		\label{eqn:d_app_lsd}
		\mathcal{L}^{app}_{LSD}=&\mathbb{E}_{\bm{x}_r \sim \mathbb{P}}\left [(D_{app}(\bm{x}_r|\bm{x}_s)-1)^2\right ] + \\ &\mathbb{E}_{\bm{x}_f \sim \mathbb{Q}}\left [(D_{app}(\bm{x}_f|\bm{x}_s)-0)^2\right ],
	\end{split}
\end{equation}
and
\begin{equation}
	\begin{split}
		\label{eqn:d_app_gp}
		\mathcal{L}^{app}_{GP}= \mathbb{E}_{\bm{x}_r \sim \mathbb{P}} [\left\| \nabla D_{app}(\bm{x}_r|\bm{x}_s)\right\|_2^2],
	\end{split}
\end{equation}
where $\bm{x}_s$ is the source frame.

Therefore, the final loss function of the discriminative part is
\begin{equation}
	\begin{split}
		\label{eqn:d_loss}
		\mathcal{L}_{D} = \mathcal{L}^{act} + \mathcal{L}^{app}.
	\end{split}
\end{equation}

For the generative part, the loss function is
\begin{equation}
	\begin{split}
		\label{eqn:g_loss}
		\mathcal{L}_{G} =  \mathcal{L}_{LSG}+\alpha\mathcal{L}_{IMG}+\beta\mathcal{L}_{FEA},
	\end{split}
\end{equation}
where $\mathcal{L}_{LSG}$ is the corresponding adversarial loss, $\mathcal{L}_{IMG}$ and $\mathcal{L}_{FEA}$ are designed for synthesizing images with better visual quality, which push the generated images towards the ground-truth images in the image space and the feature space. They are calculated by:
\begin{equation}
	\begin{split}
		\label{eqn:g_lsg}
		\mathcal{L}_{LSG}=&\mathbb{E}_{\bm{x}_f \sim \mathbb{Q}}\left [(D_{act}(\bm{x}_f|E_{gen}(\bm{r}_h,\bm{r}_v))-1)^2\right ] + \\ &\mathbb{E}_{\bm{x}_f \sim \mathbb{Q}}\left [(D_{app}(\bm{x}_f|\bm{x}_s)-1)^2\right ],
	\end{split}
\end{equation}
and
\begin{equation}
	\begin{split}
		\label{eqn:g_img}
		\mathcal{L}_{IMG}=\mathbb{E}_{\bm{x}_f \sim \mathbb{Q}, \bm{x}_r \sim \mathbb{P}}\left \|\bm{x}_f - \bm{x}_r\right \|_1,
	\end{split}
\end{equation}
\begin{equation}
	\begin{split}
		\label{eqn:g_feature}
		\mathcal{L}_{FEA}=&\sum_{i}^{K}\mathbb{E}_{\bm{x}_f \sim \mathbb{Q}, \bm{x}_r \sim \mathbb{P}}\left \|\bm{f}_{\bm{x}_f}^{i,act} - \bm{f}_{\bm{x}_r}^{i,act} \right \|_1 + \\ &\sum_{i}^{K}\mathbb{E}_{\bm{x}_f \sim \mathbb{Q}, \bm{x}_r \sim \mathbb{P}}\left \|\bm{f}_{\bm{x}_f}^{i,app} - \bm{f}_{\bm{x}_r}^{i,app} \right \|_1,
	\end{split}
\end{equation}
where $E_{gen}$ is the RF-Extractor and RNN in the generative part, $\bm{f}_{\bm{x}}^{i,act}$ refers to the feature map of $\bm{x}$ at layer $i$ in the Activity-Discriminator, $\bm{f}_{\bm{x}}^{i,app}$ refers to the feature map of $\bm{x}$ at layer $i$ in the Appearance-Discriminator, and $K$ is the total number of layers.

The whole training procedure is described in Algorithm ~\ref{alg:algo}.

\begin{algorithm}[t]
	\caption{Training algorithm for RFGAN.}
	\label{alg:algo}
	\begin{flushleft}
		\textbf{Set:} The batch size $m$ is 2, the hyperparameters $\lambda=\alpha=\beta=10.0$, the learning rate $\eta$ is 0.0002.\\
		\textbf{Initialize:} 
		Initial $\Phi_{E_{gen}}$ for the RF-Extractor and RNN in generative part,
		initial $\Phi_{E_{dis}}$ for the RF-Extractor and RNN in discriminative part,
		initial $\Phi_{G}$ for the Generator,
		initial $\Phi_{D_{act}}$ for the Activity-Discriminator,
		and initial $\Phi_{D_{app}}$ for the Appearance-Discriminator.
	\end{flushleft}
	\begin{algorithmic}[1]
		\WHILE{$\Phi_{E_{gen}}, \Phi_{G}$ has not converged}
		\STATE Sample a batch of $\{\bm{r}_h, \bm{r}_v, \bm{x}_s, \bm{x}_r\}$ from the dataset
		\STATE Update $\Phi_{E_{dis}}, \Phi_{D_{act}}, \Phi_{D_{app}}$ using Adam with:
		\STATE \quad
		$\Phi_{E_{dis}} \leftarrow \Phi_{E_{dis}} - \eta\frac{1}{m}\nabla_{\Phi_{E_{dis}}}\sum_{i=1}^{m} \mathcal{L}^{act}$
		\STATE \quad
		$\Phi_{D_{act}} \leftarrow \Phi_{D_{act}} - \eta\frac{1}{m}\nabla_{\Phi_{D_{act}}}\sum_{i=1}^{m} \mathcal{L}^{act}$
		\STATE \quad
		$\Phi_{D_{app}} \leftarrow \Phi_{D_{app}} - \eta\frac{1}{m}\nabla_{\Phi_{D_{app}}}\sum_{i=1}^{m} \mathcal{L}^{app}$
		\STATE Update $\Phi_{E_{gen}}, \Phi_{G}$ using Adam with:
		\STATE \quad
		$\Phi_{E_{gen}} \leftarrow \Phi_{E_{gen}} - \eta\frac{1}{m}\nabla_{\Phi_{E_{gen}}}\sum_{i=1}^{m} \mathcal{L}_{G}$
		\STATE \quad
		$\Phi_{G} \leftarrow \Phi_{G} - \eta\frac{1}{m}\nabla_{\Phi_{G}}\sum_{i=1}^{m} \mathcal{L}_{G}$
		\ENDWHILE
	\end{algorithmic}
\end{algorithm}

\section{Experiments}

\subsection{Implementation}
\noindent\textbf{Data} 
We collected the RF signal reflections at 20Hz from our mmWave radar system, i.e., the horizontal and vertical antenna arrays generate 20 pairs of heatmaps per second. To obtain the optical human images, we attach an RGB camera with the mmWave radar system to record videos at 10 FPS. In order to reduce the coupling between the human and the environment, we collected the data under 9 indoor scenes. There were 6 volunteers involved in the data collection and each volunteer wears multiple dresses. 

In total, we create two types of RF-Vision datasets, i.e., \textit{RF-Walk} and \textit{RF-Activity}. 
For \textit{RF-Walk}, it contains 67,860 human random walking frames and 135,720 pairs of corresponding RF heatmaps. We use 54,525 frames of human walking images and 109,050 pairs of RF heatmaps for training and the rest for testing. 
For \textit{RF-Activity}, it contains 68,680 human daily activity (e.g., stand, walk, squat, sit, etc.) frames and 137,360 pairs of corresponding RF heatmaps. We use 55,225 frames of human activity images and 110,450 pairs of RF heatmaps for training and the rest for testing. 
Each human activity frame is resized to $320\times180$, and the shape of each RF heatmap is $201\times160$. 

\noindent\textbf{Training details}
The proposed model is trained using Adam solver. The learning rate is set to 0.0002 for both the generative part and the discriminative part. The number of epochs is 80 and the batch size is 2. The hyperparameters $\lambda$, $\alpha$, and $\beta$ are equal and set to 10.0. We implement our method using PyTorch and all experiments can be run on a commodity workstation with a single GTX-1080 graphics card.

\begin{figure*}
	\begin{center}
		\includegraphics[width=\textwidth]{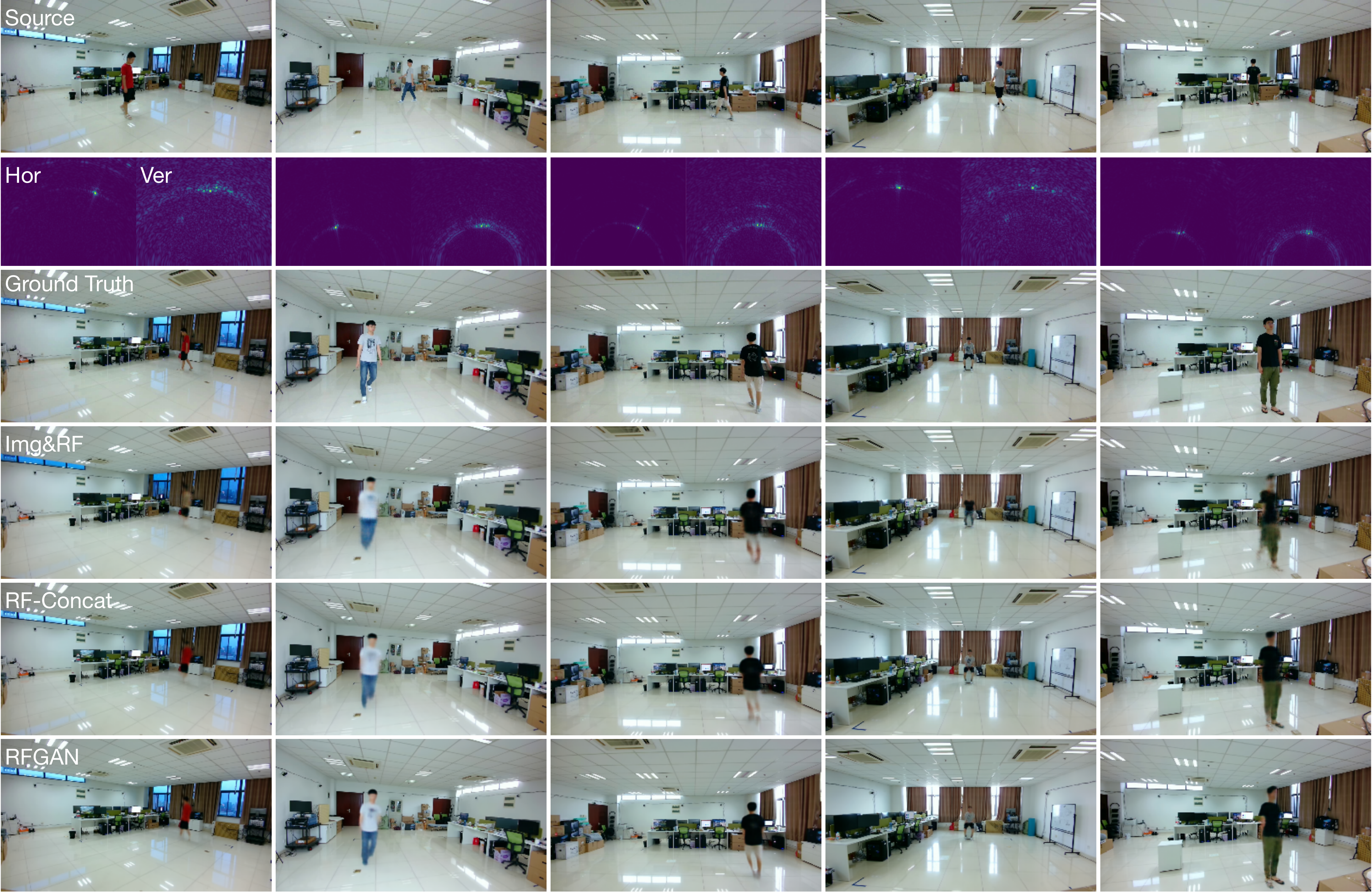}
	\end{center}
	\caption{Qualitative comparison of different methods. The 1st row shows the source frames. The 2nd row shows the horizontal and vertical RF heatmaps. The 3rd row shows the ground-truth human activity frames captured by the optical camera. The 4th to the 6th rows show the generated results by Img\&RF, RF-Concat, and RFGAN.}
	\label{fig:baselines}
\end{figure*}

\begin{table*}
	\begin{center}
		\renewcommand{\arraystretch}{1.6}
		\scalebox{1.0}{
			\setlength{\tabcolsep}{4.2mm}{
				\begin{tabular}{l|ccc|ccc}
					\hline
					& \multicolumn{3}{c|}{\textit{RF-Walk}} & \multicolumn{3}{c}{\textit{RF-Activity}} \\
					\hline
					Methods & FID $\downarrow$ & SSIM $\uparrow$ & User study $\uparrow$ & FID $\downarrow$ & SSIM $\uparrow$ & User study $\uparrow$ \\
					\hline
					\hline
					Img\&RF & 27.84 & 0.9622 & 42.11\% & 22.03 & 0.9643 & 35.89\%\\
					\hline
					RF-Concat & 21.08 & 0.9689 & 69.23\% & 19.19 & 0.9707 & 68.42\% \\
					\hline
					RFGAN & \textbf{15.75} & \textbf{0.9695} & \textbf{80.76\%} & \textbf{15.05} & \textbf{0.9708} & \textbf{78.12\%} \\
					\hline
		\end{tabular}}}
	\end{center}
	\caption{Quantitative comparison of different methods.}
	\label{tab:baselines}
\end{table*}

\subsection{Evaluation Metric}
We evaluate our proposed model from the following aspects:

\noindent\textbf{- Image Quality (FID):} 
We use the most popular metric FID~\cite{GANsTrainedByarXiv17} to evaluate the quality of the generated images. It computes the Fréchet Inception Distance between the sets of generated images and the real images. The smaller the distance, the better the quality.

\noindent\textbf{- Image Similarity (SSIM):} 
For each test sample, we calculate the visual structural similarity (SSIM)~\cite{ImageQualityAssessmentTIP04} to measure the similarity between the generated and the ground-truth human frames. A higher value means that the model can generate a human frame more similar to the ground truth.

\noindent\textbf{- User Study:} 
We conduct user surveys to evaluate whether our model can synthesize human frames with correct positions and postures. We first show our subjects some generated human frames and the corresponding ground-truth human frames, then each subject is asked to assess (yes/no) the generated results based on human positions and postures. There are 10 subjects involoved in the study. 

\begin{figure*}
	\begin{center}
		\includegraphics[width=\textwidth]{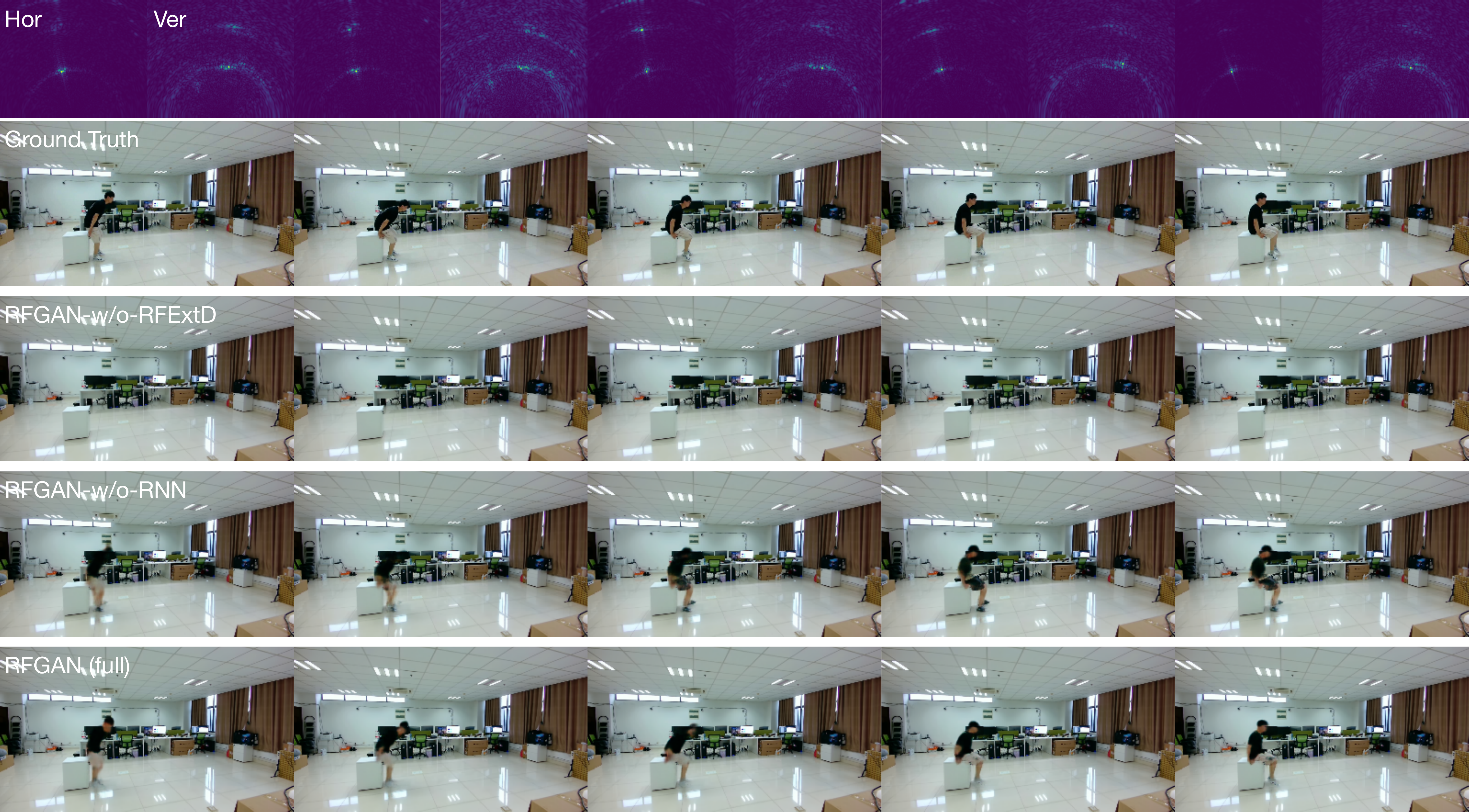}
	\end{center}
	\caption{Qualitative comparison of the ablation study. The 1st row shows the horizontal and vertical RF heatmaps. The 2nd row shows the ground-truth human activity frames captured by the optical camera. The 3rd to 5th rows show the generated results by RFGAN-w/o-RFExtD, RFGAN-w/o-RNN, and RFGAN (full).}
	\label{fig:ablation}
\end{figure*}

\subsection{Baselines}
To our knowledge, this work is the first attempt that utilizes the RF signals to generate realistic human activity frames and there is no existing and suitable baseline method to be compared with. Therefore, we modify our model with some classic approaches that are widely used in GANs or related works, and the modified models are set as the baselines:

\noindent\textbf{- Img\&RF:} 
To enable the RF-based condition setting, we propose a RF-Extractor with RNN to encode RF heatmaps and use RF-InNorm to inject the extracted information. Another alternative approach is to concatenate the RF condition with the input image directly, which is effective when the conditions have explicit guidance for GAN, e.g., pose-guided human synthesis~\cite{tang2019cycle}. However, the RF conditions are obscure data and have totally different spatial structures with optical images.

\noindent\textbf{- RF-Concat:} 
In our model, we propose a novel RF-Fusion operation to combine the horizontal and the vertical RF information, whereas the state-of-the-art approach for fusing RF information is to concatenate the features from RF signals along the channel directly, as in~\cite{zhao2018through, sengupta2020mm}. We can find most existing learning-based RF sensing works just follow the common approach in computer vision literature to combine the two-dimensional RF information. In this paper, we design a specialized operation for RF signal data.

\noindent\textbf{- Ground Truth:} 
Another baseline is the ground-truth human activity frames captured by the optical camera. 

\begin{table*}
	\begin{center}
		\renewcommand{\arraystretch}{1.6}
		\scalebox{1.0}{
			\setlength{\tabcolsep}{3.6mm}{
				\begin{tabular}{l|ccc|ccc}
					\hline
					& \multicolumn{3}{c|}{\textit{RF-Walk}} & \multicolumn{3}{c}{\textit{RF-Activity}} \\
					\hline
					Methods & FID $\downarrow$ & SSIM $\uparrow$ & User study $\uparrow$ & FID $\downarrow$ & SSIM $\uparrow$ & User study $\uparrow$ \\
					\hline
					\hline
					RFGAN-w/o-RFExtD & 58.36 & 0.9618 & 0.00\% & 45.71 & 0.9630 & 0.00\% \\
					\hline
					RFGAN-w/o-RNN & 16.41 & 0.9691 & 71.79\% & 18.11 & 0.9705 & 72.00\% \\
					\hline
					RFGAN (full) & \textbf{15.75} & \textbf{0.9695} & \textbf{80.76\%} & \textbf{15.05} & \textbf{0.9708} & \textbf{78.12\%} \\
					\hline
		\end{tabular}}}
	\end{center}
	\caption{Quantitative comparison for the ablation study.}
	\label{tab:ablation}
\end{table*}

The qualitative and quantitative comparisons are shown in Figure~\ref{fig:baselines} and Table~\ref{tab:baselines}. From the visual results, we can see that our proposed RFGAN model can capture the human position and posture information from RF signals and generate desirable activity frames. Although the baselines, i.e., RF-Concat and Img\&RF, can capture the human position information, people in the generated frames are quite blurred. The user survey results also confirm the higher position and posture accuracy of the generated human activity frames by RFGAN. According to the FID and SSIM measurements, we find that the human activity frames generated by the proposed RFGAN have better quality and are more similar to the ground truth. The experimental results demonstrate the effectiveness of our proposed RF-Fusion and the RF conditioning encoding network.

\subsection{Ablation Study}
In this subsection, we conduct ablation studies to evaluate some important components in our proposed RFGAN model:

\noindent\textbf{- RFGAN-w/o-RFExtD:} In our full RFGAN model, there are two RF-Extractors and RNNs, one in the generative part and the other in the discriminative part. In RFGAN-w/o-RFExtD, We remove the RF-Extractor and RNN in the discriminative part and use the RF fused representation extracted in the generative part as the condition for Activity-Discriminator.

\noindent\textbf{- RFGAN-w/o-RNN:} We remove the RNN module from our full RFGAN model in this setting, which means the RFGAN-w/o-RNN generates human activity frames only based on current RF signal inputs.

The qualitative and quantitative results are shown in Figure~\ref{fig:ablation} and Table~\ref{tab:ablation}. We find that the dual RF-Extractors and RNNs setting under the adversarial learning framework, i.e., RFGAN (full), can synthesize the target human activity frames, which is mainly due to the fact that the RF-Extractors and RNNs in the generative and the discriminative parts have different focuses. The RF-Extractor and RNN in the generative part pay more attention to guide the Generator for better synthesis, whereas the RF-Extractor and RNN in the discriminative part aim to help the Activity-Discriminator to distinguish different human poses. They are trained by adversarial learning.
For RFGAN-w/o-RFExtD, only one RF-Extractor and RNN are used for RF conditioning encoding. Due to the lack of supervision labels for training guidance or another RF-Extractor and RNN for adversarial learning, one RF-Extractor and RNN cannot get the desirable human activity information from the RF signals and lead to ignoring the human object in the generated frames (see 3rd row in Figure~\ref{fig:ablation}).
For RFGAN-w/o-RNN, due to the lack of information from the RF input neighbors, which can be used to adjust the current RF fused representation, it performs worse than the full RFGAN model. From the visual results, we can see the people in the generated frames by RFGAN-w/o-RNN are full of artifacts (4th row in Figure~\ref{fig:ablation}).

\subsection{Deployment in New Scene}
To deploy our model in a new scene, we do not need to retrain the whole model from the start. We can fine-tune the pre-trained RFGAN using very little data (about 40s data) to get similar results (see Table~\ref{tab:newscene}). Specifically, the learning rate during the fine-tuning process is 0.0002 for both the generative part and the discriminative part. The epochs and batch size are set to 40 and 2, respectively.  The loss functions and hyperparameters are the same with the training stage. From the quantitative results, we find that the pre-trained RFGAN model can generate desirable human activity frames in the new scene after fine-tuning with only a little data, which means our proposed model has the potential for being widely used.

\begin{table}[h]
	\begin{center}
		\renewcommand{\arraystretch}{1.6}
		\scalebox{1.0}{
			\setlength{\tabcolsep}{3.9mm}{
				\begin{tabular}{l|ccc}
					\hline
					& FID $\downarrow$ & SSIM $\uparrow$ & User study $\uparrow$ \\
					\hline
					\hline
					New Scene & 20.64 & 0.9739 & 73.33\%  \\
					\hline
		\end{tabular}}}
	\end{center}
	\caption{Quantitative results in the new scene.}
	\label{tab:newscene}
\end{table}

\begin{figure}
	\begin{center}
		\includegraphics[width=\columnwidth]{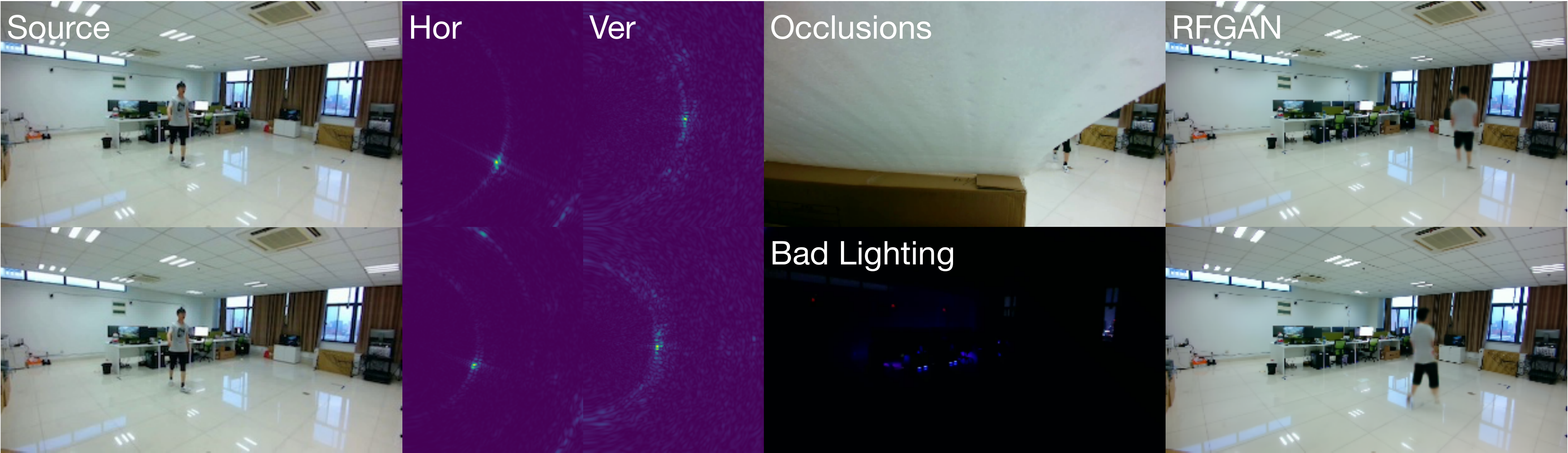}
	\end{center}
	\caption{Performance under occlusions and bad lighting.}
	\label{fig:occ_light}
\end{figure}

\subsection{Occlusions and Bad Lighting}
RF signals can traverse occlusions and do not rely on lights, thus our model can work in the occluded or bad lighting environments (see Figure~\ref{fig:occ_light}).

\section{Limitations}
Since our method relies on the natural characteristics of RF signals, the solution that we present in this paper has some limitations. Firstly, in our mmWave radar system, the depth resolution of the RF signals is about 7.5cm, and the angular resolution is about 1.3 degrees. Thus, some micro-motion behaviors that are smaller than the resolution thresholds may be missed by our model. Secondly, the operating distance of our radar system depends on the transmission power, which works up to 20m. Finally, the datasets we use in this paper mainly contain the data of human daily activities under indoor scenes. Exploring more RF-based sensing models and synthesizing people in the wild is left for future work.

\section{Conclusion}
In this paper, we aim to use RF signals to guide human synthesis. To tackle the challenge of using this new kind of driving signal, we propose a novel RFGAN model, which introduces a RF-Extractor with RNN to obtain the human activity information from the horizontal and vertical RF heatmaps and utilize the RF-InNorm to inject the information into the GAN networks. Furthermore, we propose to train the RF-Extractor and RNN under an adversarial learning framework to enable the encoding of the new kind of conditional data.
To evaluate our proposed model, we create two cross-modal datasets and the experimental results show that the RFGAN can achieve a promising performance. We believe this work opens up a research opportunity to use a new form of conditional data, i.e., RF signals, to guide the GAN model, and the performance of the RFGAN confirms that RF signals have great potential in the imaging applications.

\ifCLASSOPTIONcaptionsoff
\newpage
\fi

\bibliographystyle{IEEEtran} 
\bibliography{IEEEabrv}

\end{document}